\documentclass[10pt,prc]{article}

\usepackage{epsfig}                          
\usepackage{hyperref}

\begin{document}

\title{Self-Consistent RPA from a Coupled Cluster\\ Wave Function Perspective }

\author{ 
M. Jema\"{\i}$^{1,2}$, 
\thanks{email: jemai@ipno.in2p3.fr} 
D.S. Delion$^{3,4}$
\thanks{email: delion@theory.nipne.ro}
and 
P. Schuck$^{5,6}$ 
\thanks{email: schuck@ipno.in2p3.fr} 
}

\date{\today }
\maketitle

\begin{abstract}
\begin{center}
\parbox{14cm}{  Self-Consistent RPA is rederived in a consistent way with the help of the Coupled Cluster ground state wave function truncated at the two body level. An exact killing operator for this wave function is introduced allowing for a detailed discussion of the approximation scheme. Several exactly solvable models are reannalysed under this new perspective giving raise to a quantitative evaluation of the performances of this many body method.
}
\end{center}
\end{abstract}




\section{Introduction}
\label{Introduction}

Many Body theory is well established and practically unique on the mean field 
level. In all domains of many body physics the mean field approximation is applied in a standard way. 
When it comes to two body correlations, the approaches diverge. No standard general approach exists. 
The formalism is mostly adjusted to the problem at hand. There are the well known Quantum Monte 
Carlo methods based on a Jastrow type of ansatz \cite{Carlson,Rajagopal,BR}, the quite popular 
Coupled Cluster Theory (CCT) \cite{BR,Bishop,Walet,Bartlett}, the very successful Density Matrix Renormalisation Group method (DMRG) \cite{Hallberg}, 
and many more. In this diverse situation, it may be useful to present promising progress with yet another formalism.
In the recent past a certain category of extensions of RPA theory has been developed, 
mostly in nuclear physics, which can generically be called 'Self-Consistent 
RPA (SCRPA)' \cite{Eth,Duke,Roep,Hirsch,Storo,Jemai,Delion}. It is based on the Equation of Motion Method (EMM) initiated by D. Rowe et al. \cite{Row68}. One defines a creation operator for collective states

\begin{equation}
Q_{\nu}^{+} = \sum_{ph}(\bar X_{ph}^{\nu} a_p^{\dag} a_h-\bar Y_{ph}^{\nu} a_h^{\dag} a_p)
\label{excOp}
\end{equation}

\noindent
with $k=(p,h)$ the (particle-hole) indices corresponding to an as yet undefined
optimal single particle basis and $a^{\dag}a$ being the fermionic $ph$ pair 
operators. The excited state is then written as

\begin{equation}
|\nu\rangle = Q_{\nu}^+|0\rangle 
\label{Qpn}
\end{equation}

\noindent
where $|0\rangle$ is a postulated ground state of the system so that it is 
the vacuum to the destructor $Q_{\nu}$ for any value of $\nu$, i.e.

\begin{equation}
Q_{\nu}|0\rangle = 0~.
\label{killvac}
\end{equation}

\noindent
The amplitudes $\bar X,\bar Y$ can be obtained from the EMM outlined in \cite{Duke,Row68,RS}. The same eigen-value equations as obtained by the EMM can also be derived from the minimisation of an average 
excitation energy given by the energy weighted sum rule
\begin{equation}
\Omega_{\nu} = \frac{\langle 0|[Q_{\nu},[H,Q_{\nu}^+]]|0\rangle}
{\langle 0|[Q_{\nu},Q_{\nu}^+]|0\rangle} ~.
\label{exEnOmeganu}
\end{equation}

\noindent
where, however, the dependence of $|0\rangle $ on $\bar X,\bar Y$ in the variation is neglected.
More on this derivation of SCRPA is given in \cite{Hirsch} for the case of SCRPA in the so-called particle-particle channel, see also below. The principle is, however, the same as in the present particle-hole case. 
Replacing in this expression the correlated ground state by the Hartree-Fock 
one, leads, as well known, to the standard HF-RPA equations \cite{RS}.\\

Several remarks are in order: the replacement of the correlated ground state 
by the uncorrelated HF-one is often known in the literature under the name of
'quasi-boson approximation'. It amounts to treat the $ph$ fermion pair operators 
in $Q^+$ as ideal bosons, that is $a^+_pa_h \rightarrow B^+_{ph}$, 
with the latter the boson creation operator. 
The RPA excitation operator in (\ref{excOp}) is then identified as the 
quasi-boson operator corresponding to a 
Hartree-Fock-Bogoliubov (HFB) transformation of those bosons.
Also the killing condition of eq.(\ref{killvac}) can then be solved explicitly 
and corresponds to a HFB ground state of those 'quasi-bosons', i.e.

\begin{equation}
|0\rangle\rightarrow \exp(\frac{1}{4}\sum_{p_1p_2h_1h_2} \tilde z_{p_1h_1p_2h_2} B^+_{p_1h_1} B^+_{p_2h_2})|o)
\label{GrStBos}
\end{equation}

\noindent
with $B_{ph}|o) = 0$ and $\tilde z= \frac 12 \bar Y(\bar X)^{-1}$.

Since these things are well known, see e.g. \cite{RS}, we do not extend on them  here. 
A general fermion Hamiltonian with, e.g. a four fermion interaction can be expanded 
in a power series of bosons, as, e.g., done in the Holstein-Primakoff \cite{HP} or 
Beliaev-Zelevinsky \cite{BZ} approaches. In this way, one then arrives at a self-
consistent HFB theory for collective states of the $ph$ type, see  Sect. 9.2.3.1 of \cite{RS}.
The standard HF-RPA scheme implies a further approximation involving a 
linearisation of the HFB equations, or in other words pushing the boson expansion 
of $H$ only to quadratic order in the boson operators which is then 
diagonalised by the Bogoliubov transformation \cite{RS}. This scheme 
is usually extended to include the exchange term of the $ph$ interaction and which then 
corresponds to a linearisation of the Time-Dependent HF (TDHF) equations, i.e., standard RPA. It has 
been very successful in many domains of physics and is still widely applied at 
present.\\

Nevertheless, the quasi-boson approximation may show severe drawbacks in many 
cases and the objective of this and similar preceding works \cite{Eth,Duke,Roep,Hirsch,Storo,Jemai,Delion,Cat96}
is to get rid of this approximation, i.e. of the Pauli principle violation, as much as possible. 
The direct minimisation of the 
sum rule with respect to the amplitudes $\bar X, \bar Y$ keeping the correlated ground state can be 
performed and leads to equations which have the usual RPA structure, only the 
elements of the RPA matrix now contain single particle and two particle density 
matrices which should be evaluated with the RPA ground state 
consistent with eq.(\ref{killvac})
which, in principle, depends in a non linear way on $\bar X, \bar Y$, similar to the HFB 
approach mentioned before. Since the mathematical structure of the equations remains the 
same as in the standard case \cite{RS}, the amplitudes $\bar X,\bar Y$ obey the usual 
orthonormalisation relations and, thus, the HFB transformation (\ref{excOp}) for the 
fermion pair operators can be inverted and the $ph$ fermion pair operators 
can be expressed in terms of $Q$ and $Q^+$. In this way the specific two body 
correlation functions in the RPA matrix which contain products of two ph 
fermion pair operators can, with the help of the killing condition (\ref{killvac}), be expressed in 
terms of the $\bar X, \bar Y$ amplitudes. In the past, it remained the 
problem that the occupation numbers $n_k = \langle 0|a_k^{\dag}a_k|0\rangle$ (usually the 
single particle basis which diagonalises the s.p. density matrix is employed)
and the expectation values of the squares of the s.p. occupation number 
operator could not directly be expressed in terms of the $\bar X,\bar Y$ 
amplitudes,
see, e.g., a discussion on this point in Ref.\cite{Duke}.
Various approximate schemes were elaborated 
(see, e.g. the so-called 'Catara' and particle number approximations for $n_k$ \cite{Row68,Cat96,Cat94}) 
and then the self consistent scheme for the 
amplitudes was closed and solved numerically. This scheme was applied mostly
to model cases where one could compare with exact solutions and in most cases 
the results were excellent improving, for instance, the standard RPA results 
around phase transition points. It also could be verified in all model cases that this scheme 
solves the two particle case exactly. Therefore, this kind of SCRPA 
interpolates between the two limits which are reproduced exactly, namely 
the two body case and the dense infinite matter case. It also could be shown 
that, if the so-called 'scattering terms' in the RPA operator are included, 
the well known and appreciated qualities of standard RPA, i.e., the fullfillment 
of the energy weighted sum rule and the appearance of a Goldstone 
mode in the symmetry broken phase, are preserved in SCRPA \cite{Delion}. Inspite of 
these encouraging results, a number of open questions remained. Besides the 
just mentioned problem of how to include the occupation numbers into the self-
consistent system, there is, for instance, the question whether there exists 
a wave function which fullfills the killing condition (\ref{killvac}). 
This problem was studied in detail in an application of SCRPA in the pp-channel 
to the picket fence model 
also known as the pairing model \cite{Hirsch} in the non-superfluid phase. Indeed it is well known that RPA equations also can be written down in the pp(hh) channel, see e.g.\cite{BR,RS}. The corresponding RPA operator is given by \cite{RS}

\begin{equation}
Q_{\alpha}^+ = \frac{1}{2}[\sum_{p_1p_2} \bar X_{p_1p_2}^{\alpha} a_{p_1}^{\dag} a_{p_2}^{\dag} - 
\sum_{h_1h_2} \bar Y_{h_1h_2}^{\alpha} a_{h_1}^{\dag}a_{h_2}^{\dag}]~.
\label{Qalpha}
\end{equation}

\noindent
In the superfluid phase, the fermion operators should be replaced by BCS quasiparticle operators \cite{RS} but this shall not be considered here. The above operator is, therefore, number conserving and allows to study the excitations of the $N+2$ system as well as the ground state energy of the $N$ system, with $N$ the particle number, in the non-superfluid phase.
It can be shown that the fluctuation of the particle number remains zero in both standard and self-consistent RPA.
The particular mathematical structure of the pairing model allowed for the first time to pull 
the SCRPA scheme through without any further assumptions or approximations beyond the killing 
condition. This hinges on the fact that in this model the occupation number 
operator can exactly be expressed as a product of the two fermion pair 
operators as they appear in (\ref{Qalpha}),
and thus the system of SCRPA equations can be closed naturally. The corresponding 
results compare again excellently with the exact solution obtained from the 
Richardson equations., even for very large numbers of levels and particles. 
Nevertheless certain consistency relations following from the Pauli principle 
remained slightly violated and, indeed, it could be shown that the killing 
condition (\ref{killvac}) cannot be fullfilled with (\ref{Qalpha}) besides in the two particle case, as it must be, since the latter is 
exact within SCRPA with (\ref{Qalpha}). Before inspecting this 
killing condition in more detail in the next section, let us mention that 
there exists a further natural equation of motion which determines the 
optimal single particle basis

\begin{equation}
\langle 0|[H,Q_{\nu}]|0\rangle = 0~.
\label{miniCond}
\end{equation}

\noindent
Using in (\ref{miniCond}) the HF ground state leads back to the usual HF equations. However, 
with the correlated ground state, the single particle basis couples back to the collective excitations.
The above relation also solves the problem that the SCRPA matrix is a priori not 
symmetric because the difference in the off diagonal elements is exactly 
equivalent to this generalised mean field equation.
The objective of this paper is then that we will show how the problem of the existence of the vacuum in (\ref{killvac}) can be solved with the ground state wave function of the Coupled Cluster Theory (CCT) truncated at the two body level and, with respect to (\ref{excOp}), 
a somewhat extended RPA operator. Starting from this finding, we then will set up a consistent theory where all one and two body correlation functions involved in (\ref{exEnOmeganu}) can entirely be expressed by the RPA amplitudes $\bar X, \bar Y$ involved in (\ref{excOp}). This is paticularly true for the occupation numbers and for the two body correlation functions with either four particle or four hole indices. In this new light of the SCRPA approach, we will then review a number of model cases which in the past have been treated with the SCRPA scheme employing, however, methods for the occupation numbers and two particle densities with only $p$- or only $h$-indices which were not entirely consistent with the spirit of SCRPA. We will detail these statements in the main text. We also will give some new results when treating the various models which will highlight specific features of our approach.\\

The paper is organised as follows. In the next section, we expose the general SCRPA approach based on the CCT sub two ground state. In section \ref{Applications}, we revisit several model cases which in the past have been treated with an older less consistent version, i.e., without the use of CCT. With those examples at hand we demonstrate how the new version of SCRPA works in practice. We also investigate new relations, essentially to demonstrate the efficiency of the method. Finally in section \ref{SummDisc}, we give a summary and discuss open problems.

\section{General theory; the killing condition and the RPA ground state}
\label{GeneralTh}

As mentioned in the Introduction, the killing condition cannot be solved 
in general with the RPA excitation operator in (\ref{excOp}) or in 
(\ref{Qalpha}). 
It will be shown in this section that extending the RPA including specific two body operators allows to find a 
ground state wave function which is the vacuum to this operator. 
We, therefore, generalise the RPA operator (\ref{excOp}) in the following way
(an analogous procedure can be done with (\ref{Qalpha}), see sect. \ref{PFModel})
\begin{eqnarray}
Q_{\nu} =&& \sum_{ph} [\bar X_{ph}^{\nu} a_h^{\dag}a_p - \bar Y_{ph}^{\nu}a_p^{\dag}a_h]
\nonumber \\
&+& \frac{1}{2}\sum_{php_1p_2} \eta_{p_1p_2ph} a_{p_2}^{\dag} a_{p_1} a_p^{\dag} a_h
\nonumber \\
&-&\frac{1}{2}\sum_{phh_1h_2} \eta_{h_1h_2ph} a_{h_1}^{\dag} a_{h_2} a_p^{\dag} a_h~.
\label{exOpeta}
\end{eqnarray}

\noindent
One shows that this operator annihilates the following vacuum state (for a first account of this, see \cite{Jemai3})

\begin{equation}
|0\rangle \equiv |Z\rangle = e^Z|\mbox{HF}\rangle
\label{GstatZ}
\end{equation}

\noindent
with

\begin{equation}
Z = \frac{1}{4}\sum z_{pp'hh'} a_{p}^{\dag}a_{h}a_{p'}^{\dag}a_{h'}
\label{expZ}
\end{equation}

\noindent
where the various amplitudes are related as follows

\begin{eqnarray}
\bar Y_{ph}^{\nu}&=&\sum_{p'h'} z_{pp'hh'} \bar X_{p'h'}^{\nu} ~, 
\nonumber \\
z_{pp'hh'}&=&\sum_{\nu}\bar Y^{\nu}_{ph} (\bar X^{-1})^{\nu}_{p'h'}
\label{YZ}
\end{eqnarray}

\begin{equation}
\eta_{p_1p_2ph}^{\nu} = \sum_{h_1} z_{pp_2hh_1} \bar X_{p_1h_1}^{\nu}
\label{etapp}
\end{equation}

\begin{equation}
\eta_{h_1h_2ph}^{\nu} = \sum_{p_1}  z_{pp_1hh_2} \bar X_{p_1h_1}^{\nu}~.
\label{etahh}
\end{equation}

\noindent
The amplitude $z_{pp'hh'}$ are antisymmetric in $pp'$ and $hh'$, respectively.
With the above relations, the vacuum state is entirely expressed 
by the RPA amplitudes $\bar X,\bar Y$. We remark that this vacuum state is exactly 
the one of coupled cluster theory (CCT) truncated at the two body level \cite{BR}. 
However, the use we will make of this vacuum is very different from CCT.
Of course, for the moment, all remains 
formal because this generalised RPA operator contains, besides the standard one 
body terms, also specific two body terms which cannot be handled in a 
straightforward way. For instance, this non linear transformation cannot be 
inverted in a simple manner. One may develop approximate methods to cope with these extra terms. 
A first, simple approximation consists in replacing the occupation number operators in the $\eta$-terms in (\ref{exOpeta}) by their expectation values, that is 
$ a^{\dag}_{p_2}a_{p_1} \rightarrow \langle a^{\dag}_{p_1}a_{p_1}\rangle \delta_{p_1p_2}$ and $ a^{\dag}_{h_1}a_{h_2} \rightarrow \langle a^{\dag}_{h_1}a_{h_1}\rangle \delta_{h_1h_2}$. With the definition of the occupation numbers $ n_k = \langle a^{\dag}_ka_k\rangle$, we then obtain the following approximate form of the $Q$-operator in (\ref{exOpeta})

\begin{eqnarray}
Q_{\nu} =&& \sum_{ph} [\bar X_{ph}^{\nu} a_h^{\dag}a_p - \bar Y_{ph}^{\nu}a_p^{\dag}a_h]
\nonumber \\
&+& \frac{1}{2}\sum_{php_1} \eta_{p_1p_1ph}n_{p_1} a_p^{\dag} a_h
\nonumber \\
&-& \frac{1}{2}\sum_{phh_1} \eta_{h_1h_1ph}n_{h_1} a_p^{\dag} a_h~.
\label{exOpeta-b}
\end{eqnarray}

Evidently, this approximation, though suggestive, violates the killing condition (\ref{killvac}). However, as will be shown below with the applications, the violation remains quite moderate. On the other hand, this approximation just leads to a renormalisation of the $\bar Y$ amplitudes in (\ref{exOpeta}) and, therefore, we are back to our first ansatz of eq.(\ref{excOp}). For simplicity, we will then not change the nomenclature of the $ \bar Y$-amplitudes in the following.
We are then back to our operators in (\ref{excOp}) and (\ref{Qalpha}) where the amplitudes $(\bar X,\bar Y)$ form a complete orthonormal set of vectors as explained, e.g., in \cite{RS} and below after eq.(\ref{elemt-mat}).
This then allows to invert the operators in (\ref{excOp}) (and (\ref{Qalpha}))

\begin{equation}
a^{\dag}_p a_h = \sqrt{n_h-n_p}\sum_{\nu}[X^{\nu}_{ph}Q^{\dag}_{\nu}+Y^{\nu}_{ph}Q_{\nu}]
\label{invapah}
\end{equation}
\noindent
where we defined new amplitudes

\begin{equation}
\bar X^{\nu}_{ph} =  X^{\nu}_{ph}/\sqrt{n_h-n_p} ~,~~ 
\bar Y^{\nu}_{ph} =  Y^{\nu}_{ph}/\sqrt{n_h-n_p}
\label{XY-barXY}
\end{equation}

\noindent
so that the state $\vert \nu \rangle$ is normalised, i.e. $\langle \nu \vert \nu \rangle = \langle Z\vert [Q_{\nu},Q^{\dag}_{\nu}]\vert Z\rangle/\langle Z|Z\rangle = 1$ with

\begin{equation}
\sum_{ph}[\vert X^{\nu}_{ph}\vert ^2 - \vert Y^{\nu}_{ph}\vert ^2] = 1
\label{normCond}
\end{equation}

\noindent
which is just one of the usual RPA orthonormalisation relations of the $X,Y$ amplitudes \cite{RS} and which are used to obtain (\ref{invapah}). We, however, should always remember that the above inversion implies the approximation of the $\eta$-terms discussed above.
 
On the other hand, it is important to realise that the full RPA operator of eq.(\ref{exOpeta})
is needed in certain cases, see below, to establish the 
correct relation between $z$ and $\bar X,\bar Y$ amplitudes. It is in 
general not allowed to drop the $\eta$ terms from the beginning. It is only after the 
full set of equations has been established that 
we can approximate the $\eta$ terms for the inversion as below in (\ref{invapah}). 
The use of the CCT state $|Z\rangle$ has the great 
advantage that now in the calculation of the expectation values involved in (\ref{exEnOmeganu})  also 
the occupation numbers and their quadratic fluctuations, that is, in fact all correlation 
functions can be fully and self consistently 
incorporated into the SCRPA scheme in a natural manner. For an example, 
for the occupation numbers, one proceeds as follows

\begin{equation}
a_{h}^{\dag}a_h|Z\rangle = e^Z \tilde J_{hh} |HF\rangle
\label{mean-ahh}
\end{equation}

\noindent
with $J_{hh} = a_{h}^{\dag}a_h$ and $\tilde J_{hh} = e^{-Z}J_{hh} e^Z = J_{hh} + [J_{hh},Z]$. 
Evaluating the commutator and then using the relation

\begin{equation}
\sum_{\nu} (\bar X^{-1})_{p'h'}^{\nu}Q_{\nu} = 
a_{h'}^{\dag}a_{p'} -\sum_{ph} z_{pp'hh'} a_p^{\dag}a_h
\label{invQn}
\end{equation}

\noindent
we arrive at 

\begin{eqnarray}
n_h &=& \langle a^{\dag}_h a_h\rangle \equiv \frac{\langle Z|a_{h}^{\dag}a_h|Z\rangle}{\langle Z|Z\rangle }
\nonumber \\
&=&  1 -\frac{1}{2}\sum_{p} \langle a_{p}^{\dag} a_h a_{h}^{\dag} a_{p}\rangle ~.
\label{valzahahz}
\end{eqnarray}

\noindent
For the evaluation of the two body term in (\ref{valzahahz}), we will use the inversion (\ref{invapah}) of the $Q$-operators and obtain

\begin{equation}
n_h \equiv \langle a_{h}^{\dag}a_h\rangle = 1-\frac{1}{2}\sum_{p,\nu} (n_h-n_{p}) |Y^{\nu}_{ph}|^2~.
\label{exp-apah}
\end{equation}

The same can be repeated for $n_p$ leading to a linear system of equations for $n_h,n_p$ which can be solved. The quadratic occupation number fluctuations yield slightly complicated expressions as we will see in the next sections where applications are presented.\\

In summary, we have constructed a fully self consistent RPA scheme based on the vacuum state $|Z\rangle$ which corresponds to the CCT sub two ansatz. 
Indeed, using the EMM of Rowe \cite{Row68} or the minimisation of (\ref{exEnOmeganu}), we arrive at the standard RPA equations for the $X$ and $Y$ amplitudes, i.e.

\begin{equation}
\left( 
\begin{array}{cc}
 A &  B \\ 
-B & -A
\end{array}
\right) \left( 
\begin{array}{c}
X^{\nu } \\ 
Y^{\nu }
\end{array}
\right) =\Omega _{\nu}
\left( 
\begin{array}{c}
X^{\nu } \\ 
Y^{\nu }
\end{array}
\right) 
\label{RPAeqGen}
\end{equation}
\noindent
with
\begin{eqnarray}
A_{ph,p'h'} &=&  \frac{\langle [a^{\dag}_ha_p,[H,a^{\dag}_{p'}a_{h'}]] \rangle }{ \sqrt{n_h-n_p}\sqrt{n_{h'}-n_{p'}}} ~,
\nonumber \\ 
B_{ph,p'h'} &=& -\frac{\langle [a^{\dag}_ha_p,[H,a^{\dag}_{h'}a_{p'}]] \rangle }{ \sqrt{n_h-n_p}\sqrt{n_{h'}-n_{p'}}}
\label{elemt-mat}
\end{eqnarray}

\noindent
where $\langle ...\rangle = \langle Z|...|Z\rangle/\langle Z|Z\rangle$ as in (\ref{valzahahz}). The equations (\ref{RPAeqGen}) have exactly the same mathematical structure as the standard RPA ones. Therefore the $X,Y$ amplitudes obey to the orthonormalisation conditions as, e.g., explained in \cite{RS}. It is easy to convince oneself that the commutators in (\ref{elemt-mat}) lead to one and two body correlation functions only, if a Hamiltonian with a two body interaction is used (plus three body correlation functions with a three body force). With our wave function, those correlation functions can now be expressed
with (\ref{killvac})
 entirely by the amplitudes $X$ and $Y$
leading to a fully closed system of equations.
This is a big step beyond standard RPA where the correlation 
functions in above equation are evaluated with the HF ground state. 
In a way, this SCRPA scheme now corresponds to a full evaluation of the Bogoliubov transformation (\ref{excOp}) of 
fermion pair (ph) operators. As will be shown in the next section, the 
numerical results with this scheme are excellent for various non trivial 
model cases. However, due to the approximation of the $\eta$-terms, the whole 
scheme is not entirely Raleigh-Ritz variational and slight violations of the Pauli principle are still inherent. These errors are unavoidable in a self consistent theory for two body correlations based on a coherent state. Within SCRPA they remain, however, small as will be discussed later.

\section{Applications}
\label{Applications}

In this section, we will revisit several model cases which in the past have been treated with SCRPA, however, without the use of CCT and, therefore, with a less consistent formalism. We only will consider the symmetry unbroken phases of those models and, thus, the calculations can not be applied much beyond the transition point where standard RPA breaks down. This should be kept in mind when considering the results in the following.

\subsection{The two level Lipkin model}
\label{Lipkin2}

A common testing ground of many body theories has been the two level Lipkin 
model, see, e.g. \cite{RS}. Its Hamiltonian is of the following form

\begin{equation}
H=\varepsilon J_0 -\frac{V}{2}[J_+J_+ + J_-J_-]
\label{Ham-Li2}
\end{equation}

\noindent
with $J_0 = \frac{1}{2}\sum_m[a_{1m}^{\dag}a_{1m} - a_{0m}^{\dag}a_{0m}]$, $ J_+ = 
\sum a_{1m}^{\dag}a_{0m}$, $J_-= J_+^{\dag}$. For further details, the reader is 
referred to the literature \cite{RS}. Logically, the $Z$ operator of the 
vacuum (\ref{exOpeta}) state is given by

\begin{equation}
Z=zJ_+J_+ \label{Z-Li2}
\end{equation}

\noindent
with $z=\bar Y/(N\bar X)$ and the generalised RPA detruction operator $Q=\bar XJ_- - \bar Y(1-\eta J_0)J_+$ 
with $\eta = 2/N$. Let us 
demonstrate that the $\eta$ term in $Q$ has to be kept in setting up the equations 
in order to get the correct expression for $z(\bar X,\bar Y)$. Indeed, 
the killing $Q|Z\rangle = 0$ condition leads with $|Z\rangle = e^Z|\mbox{HF}\rangle$ to 

\begin{eqnarray}
&&2\bar Xz-\eta \bar Y=0~,
\nonumber \\
&&2\bar Xz(N-1)-\frac{1}{2}\eta \bar Y(N-2)-\bar Y=0~.
\label{Rel-XYZ}
\end{eqnarray}

\noindent
One sees that neglecting in these two eqs the $\eta$ terms, leads to a wrong result for $z$.
So after having established the full theory, we now will neglect  
the $\eta$ term for the inversion as in (\ref{invapah}) and apply our scheme as outlined in the preceding section. 
One obtains for the SCRPA matrix elements given in (\ref{RPAeqGen})  

\begin{figure}[ht]
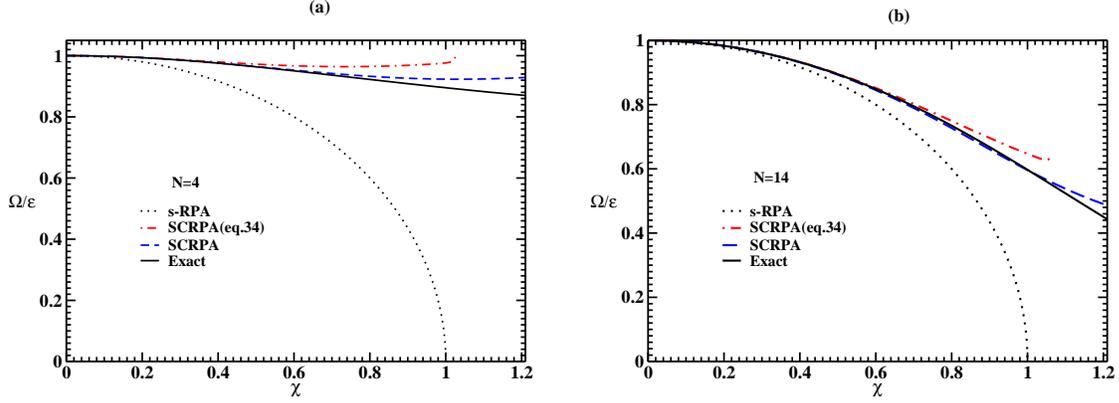

 \begin{center}
  \epsfig{figure=Li-N4.eps,width=7cm}
  \hspace{0.5cm}
  \epsfig{figure=Li-N14.eps,width=7cm}
\caption{"(Color online)" Energy of first excited state with standard RPA, SCRPA(eq.\ref{VJ-Li2-Catara}) 
with the density in eq.(\ref{VJ-Li2-Catara}), and SCRPA compared to exact 
solutions for $N=4$ and $N=14$ as function of interaction $\chi = V(N-1)/\varepsilon $. }
\label{Li2-N4}
 \end{center}
\end{figure}

\begin{eqnarray}
A &=& 1+2V\frac{\langle J_{+}^{2}\rangle }{\langle -2J_{0}\rangle } ~,~
\nonumber \\
B &=& V\left[ 1-\frac{\langle (2J_{0})^{2}\rangle -2\langle J_{+} J_{-}\rangle }{\langle -2
J_{0}\rangle }\right]~.
\label{ABLip2}
\end{eqnarray}
With the techniques outlined in (\ref{mean-ahh})-(\ref{valzahahz}), we obtain
\begin{eqnarray}
\langle J_0\rangle &=& -\frac{N}{2} + \frac{2}{N} \langle J_+ J_-\rangle ~~~~~~~~\mbox{and}
\nonumber \\
\langle J_+ J_-\rangle &=& -2\langle J_0\rangle Y^2 = \frac{NY^2}{1+\frac{4}{N}Y^2}
\end{eqnarray}
\noindent
with

\begin{eqnarray}
\langle J_0\rangle &=& -\frac{N/2}{1+\frac{4}{N}Y^2}~, ~~~~~~~~~~\mbox{and}
\nonumber \\
\langle J_{+}^{2}\rangle &=& XY\langle -2 J_{0}\rangle 
\nonumber \\
&=& XY \left[ N-4\frac{Y}{NX}\langle J_{+}^{2}\rangle \right] =\frac{NXY}{1+\frac{4}{N} Y^{2}}~.
\label{meanValJ0Li2}
\end{eqnarray}

\noindent
In the expression (\ref{ABLip2}), we also have to calculate $\langle \hat{J}_{0}^{2}\rangle $. 

\noindent
We can use 
\begin{eqnarray}
J_0\vert Z\rangle &=&\left(-\frac{N}{2} +2zJ^{2}_{+}\right)\vert Z\rangle ~,~~~~~~~~~\mbox{and thus}
\nonumber \\
J_0J_0\vert Z\rangle &= & \left(-\frac{N}{2}J_0 +2z J_0 J^{2}_{+}\right)\vert Z\rangle
\nonumber \\
 &=&\left(\frac{N^2}{4} -2z(N-2) J^{2}_{+} +4z^2 J^{4}_{+}\right)\vert Z\rangle ~.
 \label{demstLi2-1}
\end{eqnarray}

\noindent
We can express the action of $ J_{+}^{4}\vert Z\rangle $ as
\begin{eqnarray}
Nz J_{+}^{2} \vert Z\rangle &=&  J_{+} J_{-}\vert Z\rangle~,
\nonumber \\
 &=& J_{+} \left( 2z(N-1)  J_{+} -4z^2 J^{3}_{+}\right) \vert Z\rangle
\nonumber \\
 &=& \frac{4N}{N-2} z^2 J^{4}_{+} \vert Z\rangle 
 \label{demstLi2-2}
\end{eqnarray}

\noindent
what implies that

\begin{eqnarray}
Nz\langle J_{+}^{2}\rangle &=& \langle J_{+} J_{-}\rangle  ~,
\nonumber \\
\mbox{and}~~~~~~~~~~~~~~~~~~~
4z^{2}\langle J_{+}^{4}\rangle &=& \left( 1-\frac{2}{N}\right) \langle J_{+} J_{-}\rangle
\label{demstLi2-3}
\end{eqnarray}
Therefore
\begin{equation}
\langle J_{0}^{2}\rangle = \frac{N^{2}}{4}-\frac{1}{N}\left( N-2\right) 
\langle \hat{J}_{+}\hat{J}_{-}\rangle ~.
\label{meanValJ02-Li2}
\end{equation}

\noindent
We check that $\langle \hat{J}_{0}^{2}\rangle = 1$ for $N=2$, which represents the exact result.

\noindent
For the elements of the RPA matrix this gives
\begin{eqnarray}
A &=& 1+2VXY ~,
\nonumber \\
B &=& V\left[ 1-N +2(1-4/N)Y^2\right]~.
\label{ABLip2XY}
\end{eqnarray}

\begin{figure}[ht]
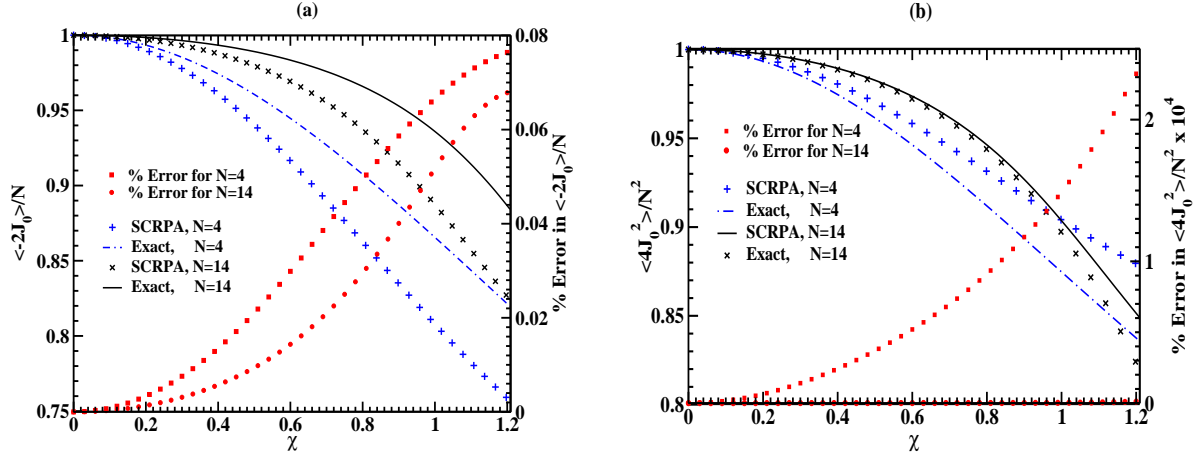

\begin{center}
  \epsfig{figure=ERR_J0.eps,width=7.5cm,height=6cm}
  \hspace{0.5cm}
  \epsfig{figure=ERR_J02.eps,width=7.5cm,height=6cm}
\caption{"(Color online)" The percentage error between SCRPA and exact solution for $N=4$ and $N=14$ with $\% \mbox{Error} =|\langle ...\rangle_{exact} - \langle ...\rangle_{SCRPA}|/\langle ...\rangle_{exact} $  as function of interaction $\chi = V(N-1)/\varepsilon $ (right ordinate). Also the mean values of $\langle -2J_0\rangle/N $ and $\langle 4J^2_0\rangle/N^2$ as as function of interaction $\chi$ are shown (left ordinate). Notice that the scale on left ordinates only covers the range from 0.75 to 1.0 (upper panel) and 0.80 to 1.0 (lower panel). }
\label{ERR_J0}
\end{center}
\end{figure}

\noindent
It is interesting to compare  results with the expression used in earlier 
SCRPA schemes based on a pure boson approximation, i.e., $Q^{\dag} = XB^{\dag}-YB$ and $B^{\dag} = XQ^{\dag} + YQ$ with $\langle J_0\rangle = -N/2 +\langle B^{\dag}B\rangle = -N/2 + Y^2$ or resummed as in \cite{Cat94}

\begin{equation}
\langle J_0 \rangle = - \frac{N/2}{1+\frac{2}{N} Y^2}
\label{VJ-Li2-Catara}
\end{equation}

\noindent
and a more complicated expression for $\langle J_0J_0 \rangle$. 
We see that the bosonisation does not yield expressions equal to our result based on a fermion pair algebra. For the term  lowest order in $Y^2$, the bosonisation yields a factor of two lower than our present expression. Still another factor of two lower gives the so-called Catara expression \cite{Cat96}.
In Fig.\ref{Li2-N4}, we compare the SCRPA results for the excitation energy with the two versions of $\langle J_0\rangle$. We see that in general, the results are very good in comparison with the exact ones, however, the expression (\ref{demstLi2-2}) obtained with our CCT wave function yields slightly better ones. 
In Fig.\ref{ERR_J0} we give the relative error for the occupation factors and their fluctuations. Again, the results with a maximum error of about eight percent are rather good. 

\begin{figure}[th]
  \begin{center}
    \leavevmode
     \epsfig{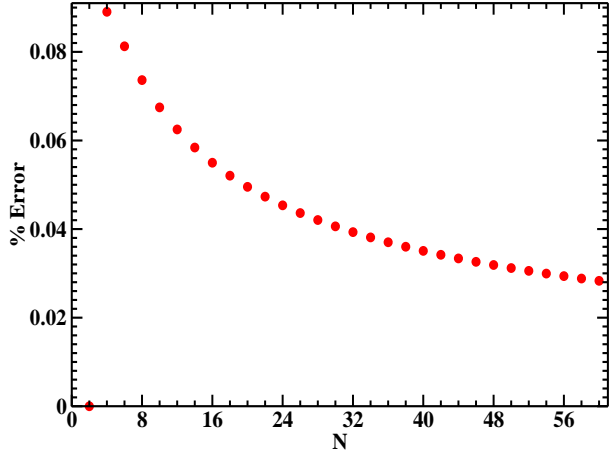}
      \caption{ The percentage error between the SCRPA and exact solution for $\chi=1$  with $r=-\sqrt{\langle J_0^2\rangle}/\langle J_0\rangle $ and $\% \mbox{Error} =|r_{exact} - r_{SCRPA}|/r_{exact} $  as a function of  $N $. Notice the zero error for $N=2$.}
  \label{ERR_N}
 \end{center}
 \end{figure}

In Fig.\ref{ERR_N}, we display the relative error for the ratio $r = \sqrt{\langle J_0^2\rangle}/\langle -J_0\rangle$ as a function of particle number $N$ for the coupling $\chi=1$. We remark that for $N=2$ the result is, as expected, exact. However, immediately after, for $N=4$, the error jumps to its maximal value of about nine percent, before decreasing again. For $N \rightarrow \infty$ the result will be exact, since RPA becomes exact in this limit for the Lipkin model. Therefore, SCRPA interpolates between those two exact limits. For the following it should be kept in mind that the cases with a small number of particles represent the most severe tests of the theory.\\

\subsection{Three Level Lipkin Model}
\label{Lipkin3}

In the three level Lipkin model, one considers the following Hamiltonian

\begin{equation}
H=\sum_{k=0}^2 \epsilon_k J_{kk} -\frac{V}{2}\sum_{p=1}^2(J_{p0}J_{p0} + J_{0p}J_{0p})
\label{Ham-Lip-3}
\end{equation}

\noindent
where $J_{kk} = a^{\dagger}_{k} a_{k}, ~ J_{p0} =  a^{\dagger}_{p} a_{0} $.
From the three single particle energies $\epsilon_k$, we choose $\epsilon_0$ 
to be below the Fermi level and the other two above. The model has the 
interesting feature that if the two upper levels are degenerate, then in 
the standard HF-RPA scheme, in the symmetry broken phase, a Goldstone mode 
appears, that is, there exists a continuously broken symmetry. However, 
again as in the other models we only will consider here the 'spherical' case.
As before, we introduce the vacuum $|Z\rangle = e^Z|HF\rangle$ with

\begin{equation}
Z=\frac{1}{2}\sum_{p_1p_2} z_{p_1p_2}J_{p_10}J_{p_20} ~.
\label{ZLi3}
\end{equation}

\noindent
In analogy to the preceding sections, one can show that the following RPA 
destruction operator kills this vacuum

\begin{equation}
Q_{\nu}= \sum_p[\bar X_p^{\nu}J_{0p} - \bar Y_p^{\nu}J_{p0} + \eta_{0p}^{\nu}J_{00}J_{p0}]
\label{QnLi3}
\end{equation}

\noindent
with

\begin{equation}
\eta_{0p}^{\nu} = \sum_{p'}\bar X_{p'}^{\nu}z_{p'p}~;~~
(N-1)z_{pp'}=\sum_{\nu}(\bar X^{-1})_p^{\nu}\bar Y_{p'}^{\nu}~.
\label{Rel-etaXYZLi3}
\end{equation}

\noindent
Also the occupation numbers can be calculated, similar to the procedure in the preceding section for the two level Lipkin model

\begin{eqnarray}
n_0 &=& \langle J_{00}\rangle \equiv \frac{\langle Z|J_{00}|Z\rangle}{\langle Z|Z\rangle} 
\nonumber \\
&=& N-\frac{1}{(N-1)}\sum_{p} \frac{\langle Z|J_{p0}J_{0p}|Z\rangle}{\langle Z|Z\rangle}
\nonumber \\
&=& N - \frac{1}{(N-1)}\sum_{p,\nu} |\bar Y_{p}^{\nu}|^2 
\nonumber \\
&=& N-\sum_p(n_0-n_p)S_p
\label{n0Li3}
\end{eqnarray}
\begin{eqnarray}
n_p &=& \frac{1}{(N-1)}\frac{\langle Z|J_{p0}J_{0p}|Z\rangle}{\langle Z|Z\rangle}
\nonumber \\
&=&\frac{1}{(N-1)}\sum_{\nu} |\bar Y_{p}^{\nu}|^2 
\nonumber \\
&=& (n_0-n_p)S_p
\label{npLi3}
\end{eqnarray}
with  $S_p= \frac{1}{(N-1)}\sum_{\nu} |Y_{p}^{\nu}|^2 $. This system of equations for $n_0$ and $n_p$ can easily be solved. The results (not shown) are of similar quality as in the case of the two level Lipkin model of preceding section.\\

\noindent
In a similar way the quadratic occupation numbers can be calculated. We obtain 
the following expressions. 

\begin{eqnarray}
\langle J_{00}J_{00}\rangle = N^2 &-& \frac{2N}{N-1} \sum_p \langle J_{p0}J_{0p}\rangle 
\nonumber \\
&+& \frac{1}{(N-1)^2}\sum_{pp'}\langle J_{p0}J_{0p} J_{p'0}J_{0p'}\rangle
\nonumber \\
\langle J_{00}J_{p'p'}\rangle = && \frac{N}{(N-1)} \langle J_{p0}J_{0p}\rangle 
\nonumber \\
&-& \frac{1}{(N-1)^2} \sum_p\langle  J_{p0}J_{0p} J_{p'0}J_{0p'}\rangle
\nonumber \\
\langle J_{pp}J_{p'p'}\rangle = && \frac{1}{(N-1)^2} \langle J_{p0}J_{0p}J_{p'0}J_{0p'}\rangle ~.
\label{J00JppLi3}
\end{eqnarray}

\begin{figure}[ht]
\begin{center}
    \leavevmode
     \epsfig{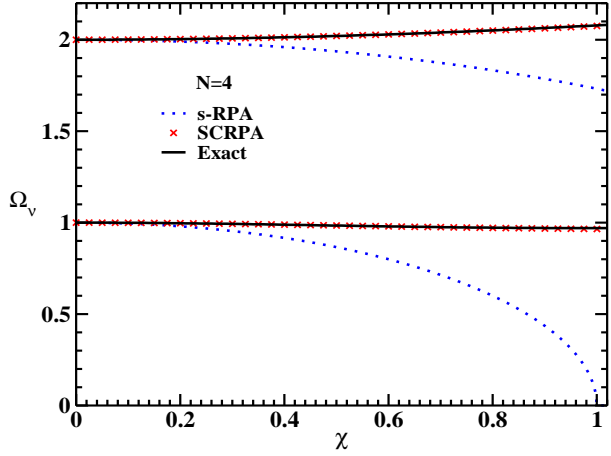}
      \caption{"(Color online)" The first and second excited state for the 3-level 
      Lipkin model with SCRPA and exact solution for $N=4$  as 
      function of interaction $\chi = V(N-1)/\varepsilon $.}
  \label{ENERG-N4-L3}
 \end{center}
\end{figure}

\noindent
Given the expressions in (\ref{J00JppLi3}), we can again use the inversion (\ref{invapah})
of the RPA operator and obtain the fluctuation of the occupation numbers as a function of the RPA amplitudes. However, generally, one will factorise the four body correlation functions in antisymmetrised products of two body correlation functions. This will maintain the property that the two particle case is solved exactly. We will not go further into the details of these procedures here, since it goes in complete analogy to what has been done for the pairing model in \cite{Hirsch} which will be shortly revisited in the next subsection. Having all correlation functions at hand, we can calculate the excitation energies. They are displayed in Fig.\ref{ENERG-N4-L3}. We see that the results are again very good, the maximum error at $\chi=1$ being  about $0.5$ percent.

\subsection{The picket fence ( pairing) model revisited}
\label{PFModel}

As already mentioned, the picket fence model (PFM) is the only one where in  the past the 
SCRPA scheme could be pulled through for the first time without any approximation and without the 
explicit knowledge of the vacuum \cite{Hirsch}. This stemmed from the fact that in this 
particular model with the Hamiltonian

\begin{equation} 
H= \sum_i \varepsilon_i N_i  + V\sum_{ik} P^+_i P_k
\label{Ham-PFM}
\end{equation}

\noindent
and N two fold degenerate and equidistant levels, the 
occupation number operators can exactly be expressed by the product of two 
fermion pair operators, that is

\begin{equation}
N_i=2P_i^{\dag}P_i
\label{Numb-PFM}
\end{equation}

\noindent
with $N_i = a_{i+}^{\dag}a_{i+} + a_{i-}^{\dag}a_{i-}$ and $P_i^{\dag} = 
a_{i+}^{\dag}a_{i-}^{\dag}$. It is seen that the pair operators are the ones which 
enter the Bogoliubov transformation of fermion pairs in the pp-SCRPA as 
exemplified in eq.(\ref{Qalpha}) and, therefore, with (\ref{killvac}) 
it was possible in \cite{Hirsch} to 
calculate $\langle N_i\rangle$ and $\langle N_iN_j\rangle$ completely 
selfconsistently and without the use of any procedures external to the SCRPA 
ones. We also remark that the evaluation of $\langle N_iN_j\rangle$ 
necessitates the knowledge of four particle correlation functions what makes 
the approach slightly heavy. However, factorisation $\langle N_iN_j\rangle 
\sim \langle N_i\rangle \langle N_j\rangle$ turned out to work quite well, 
thus simplifying the expressions substantially \cite{Duke-PLB}.\\

How is all this linked to our wave function approach? Let us calculate the 
occupation numbers $\langle N_i\rangle$ from our vacuum wave function. In the pp-
case it writes

\begin{equation}
|Z\rangle = \exp(-\sum_{ph} z_{ph} P_p^{\dag} P_h )|HF\rangle
\label{GstatZ-PFM}
\end{equation}

\noindent
with $p,h$ indices again indicating single particle states above, below the Fermi level, 
respectively.

\noindent
It can be shown that the following destruction operator kills the above vacuum

\begin{equation}
Q_{\alpha} = \sum_p \bar X_p^{\alpha}P_p + \sum_h \bar Y_h^{\alpha} P_h - \sum_{ph} \eta_{ph}^{\alpha} N_pP_h
\label{Qalpha-PFM}
\end{equation}

\noindent
where

\[
z_{ph} = \sum_{\alpha} (\bar X^{-1})^{\alpha}_p \bar Y_h^{\alpha}
\]
\[
2\eta_{ph}^{\alpha} = - \bar X_p^{\alpha} z_{ph}~.
\]

\noindent
We calculate the occupation numbers as in the $ph$ case above. For example

\begin{eqnarray}
N_p|Z\rangle &=& e^Z \tilde N_p|HF\rangle~, ~~~~\tilde N_p=e^{-Z}N_pe^Z
\nonumber \\
&=& -2\sum_h z_{ph}P_p^{\dag}P_h|Z\rangle ~.
\label{expNp-PFM}
\end{eqnarray}

\noindent
As in the general case, we check the identity $P_p|Z\rangle = -\sum_h z_{ph}P_h|Z\rangle$. With this,
we see that we get exactly the same expression for the occupation numbers as in \cite{Hirsch} and
discussed above without the use of the vacuum, namely $\langle N_i\rangle = 
2\langle P_i^{\dag}P_i\rangle$. One shows that this also holds for the squares 
$\langle N_iN_j\rangle $. Therefore, our published results in \cite{Hirsch} are fully 
equivalent to our scheme advocated above using the vacuum $|Z\rangle$. 
Commenting on the SCRPA results which can be found in \cite{Hirsch}, we see that they 
are very good. On the other hand, we also see from the sum rule relation (69), i.e. $\sum_{pp'}\langle N_pN_{p'}\rangle = \sum_{ph}\langle (2-N_h)N_p\rangle$ and Tables VII and XI in \cite{Hirsch} that the sum rule, i.e., the 
Pauli principle, is still slightly violated, of the order of $4-5$ percent, what stems from the (mild) 
violation of the killing condition $Q|Z\rangle =0$ using the restricted form 
of the pp-RPA operator when the $\eta$ term in (23) is neglected. 
Let us mention that SCRPA was solved in \cite{Hirsch}
among others for the case of 100 levels where it is even difficult to solve the 
problem with the Richardson equations. Again the model was treated only in the 
symmetry unbroken phase and we will discuss about this later. A very instructive example is the $N=2$ case. Though already presented in \cite{Hirsch}, let us comment here again. In standard RPA the excitation energy is given by

\[ E \propto \sqrt{1-G}\]

\noindent
whereas in SCRPA the result is

\[ E \propto \sqrt{1+G}~.\]

\noindent
The latter coincides, as already mentioned, with the exact result. The RPA result shows the usual BCS instability at $G=1$. With SCRPA the vertex renormalisation from the self consistency, i.e. screening, has effectively turned the sign of G around and with screening the effective interaction is now {\it repulsive}! This stems from the fact that for $N=2$ the constraint from the Pauli principle is, as one easily realises, of maximum importance. SCRPA testifies again with this example that the Pauli principle is very well respected.

Let us also mention that the SCRPA scheme has been generalised to finite 
temperatures in an application to the PFM in \cite{Storo} with the same quality of results as at zero temperature. In particular it could be shown that also in the PFM, there opens a 
pseudo gap in the level density  approaching the critical temperature from above.

\subsection{The Hubbard Model}
\label{HubbardMod}

One further application of SCRPA concerned in the past the 1D Hubbard model 
with periodic boundary conditions \cite{Jemai}. The two site case with half filling, 
i.e. the two particle case (the so-called Hubbard molecule) turned out, as expected, to be solved exactly again, see in this respect \cite{Seibold,Myake}. In this 
model, like for the PFM, the knowledge of the vacuum is not necessary to close 
the SCRPA eqs in a natural way. 

In \cite{Jemai} SCRPA was solved for each $q$-value separately in order to avoid problems with the implicit 
channel coupling via the non-linearities where $q'$-values different from the 
external one ($q$) can appear in the correlation functions. We will come back to this point in the last section of this paper.

For completeness, we shortly outline how our formalism works in the Hubbard 
model. In momentum space, the Hamiltonian is given by

\begin{eqnarray}
\label{Ham-HubbMod}
H=&&\sum_{{\bf k},\sigma} (\epsilon_k - \mu)\hat n_{{\bf k}, \sigma}
\\
&&+\frac{U}{2N}
\sum_{{\bf k},{\bf p},{\bf q}, \sigma} a^{\dag}_{{\bf k},\sigma}a^{\dag}_{{\bf k}+{\bf q},
\sigma} a_{{\bf p}, -\sigma}a_{{\bf p}-{\bf q}, -\sigma}
\nonumber
\end{eqnarray}

\noindent
where $\hat n_{{\bf k},\sigma}=a^{\dag}_{{\bf k},\sigma}a_{{\bf k},\sigma}$ is the 
occupation number operator and the single particle energies are given 
by $\epsilon_{{\bf k}} = -2t\sum_{d=1}^D cos(k_d)$ with the lattice spacing 
set to unity. It is convenient to transform the creation and annihilation 
operators $a^{\dag},a$ to HF quasi-particle operators. In 1D, we have

\begin{equation}
a_{h,\sigma} = b^{\dag}_{h, \sigma}~,~~~~a_{p,\sigma} = b_{p,\sigma}
\label{transf-atobHubb}
\end{equation}

\noindent
where $h$ and $p$ are momenta below and above the Fermi momentum, 
respectively, so that $b_{k,\sigma}|\mbox{HF}\rangle =0$ for all $k$ where 
$|\mbox{HF}\rangle$ is the Hartree-Fock ground state in the plane wave basis.
Introducing the operators

\begin{equation}
\tilde n_{k,\sigma} = b^{\dag}_{k,\sigma}b_{k,\sigma} ~,
\label{occpNumHubb}
\end{equation}

\begin{equation}
J^-_{ph,\sigma}=b_{h,\sigma}b_{p,\sigma}~,~~~~J^+_{ph,\sigma}=(J^-_{ph,\sigma})^{\dag}
\label{quasiPartOp}
\end{equation}

\begin{figure}[ht]
  \begin{center}
    \leavevmode
     \epsfig{file=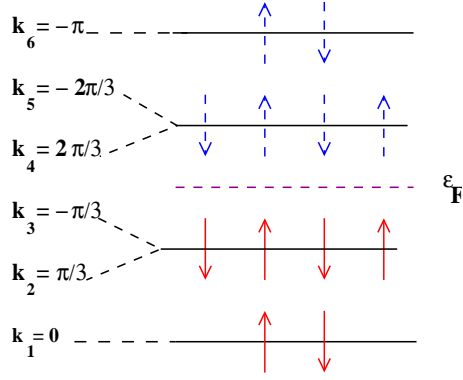,width=6cm,height=5cm}
      \caption{"(Color online)" 
Excitation spectrum of the HF ground state $U=0$ for the
chain with six sites at half filling and projection of spin $m_s=0$. 
The occupied states are represented by the solid arrows and those not 
occupied are represented by the dashed arrows. 
}
  \label{Hub3}
 \end{center}
\end{figure}

\noindent
we write for the vacuum $|Z\rangle = e^Z|\mbox{HF}\rangle$ with

\begin{equation}
Z=\sum_{p_1p_2h_1h_2} z_{p_1p_2h_1h_2}J^+_{p_1h_1,+}J^+_{p_2h_2,-} ~.
\label{ZHubb1}
\end{equation}

\noindent
It can be shown that the following operator 

\begin{eqnarray}
Q_{q,\nu}=\sum_{ph,\sigma}[ && \bar X^{\nu}_{ph,\sigma}J^-_{ph,\sigma} - \bar Y^{\nu}_{ph, -\sigma}
J^+_{ph, -\sigma}] 
\nonumber \\
&+& \frac{1}{2}\sum_{hh_1p_2h_2}\eta^{\nu}_{hh_1p_2h_2} b^{\dag}_{h_1,\sigma}b_{h,\sigma}J^+_{p_2h_2,-\sigma} 
\nonumber \\
&+&\frac{1}{2}\sum_{pp_1p_2h_2} \eta^{\nu}_{pp_1p_2h_2}b^{\dag}_{p_1,\sigma}b_{p,\sigma}J^+_{p_2h_2,-\sigma}
\label{exOpHubb}
\end{eqnarray}

\noindent
with 

\begin{eqnarray}
\eta^{\nu}_{hh_1p_2h_2} = \sum_p \bar X^{\nu}_{ph,\sigma}z_{pp_2h_1h_2}~,
\nonumber \\
\eta^{\nu}_{pp_1p_2h_2} = \sum_h\bar X^{\nu}_{ph,\sigma}z_{p_1p_2hh_2}~,
\label{eta-Hubb}
\end{eqnarray}

\noindent
and

\begin{equation}
z_{pp'hh'} = \sum_{\nu}(\bar X^{-1})^{\nu}_{ph,\sigma}\bar Y^{\nu}_{p'h',-\sigma}~,
\label{expres-ZHubb}
\end{equation}

\noindent
annihilates this vacuum. We see that for the Hubbard model, the formalism coincides 
practically with the general one. The change in sign for the $\eta$ amplitudes 
stems from our transformation to HF  operators. 
In Fig.\ref{Hub3}, we present for completeness the level scheme for the six site case.
With our usual technique for the evaluation of the occupation numbers, we also verify 
straightforwardly that, e.g.

\begin{figure}[ht]
  \begin{center}
    \leavevmode
     \epsfig{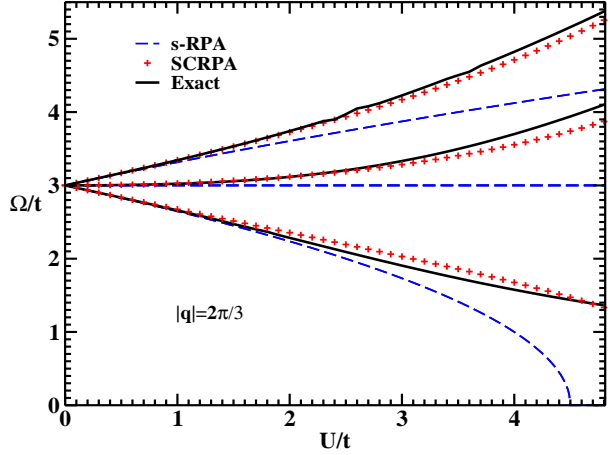}
      \caption{"(Color online)" 
      Energies of excited states in the standard RPA, SCRPA,
and exact cases as a function of $U/t$ for six sites with spin projection
$m_{s} = 0$ and for $|q|=2\pi/3$. 
}
  \label{Rep2p3}
 \end{center}
\end{figure}

\begin{eqnarray}
\langle n_{p\sigma}\rangle = \langle \tilde{n}_{p\sigma}\rangle &=& \sum_{h}\langle J^{+}_{ph\sigma} J^{-}_{ph\sigma}\rangle
\nonumber \\
& = & \sum_{h,\nu} (1-\langle M_{ph\sigma}\rangle) |Y^\nu_{ph\sigma}|^2, 
\nonumber \\
\langle n_{h\sigma}\rangle =1- \langle \tilde{n}_{h\sigma}\rangle & = & 1- \sum_{p}\langle J^{+}_{ph\sigma} J^{-}_{ph\sigma}\rangle
\nonumber \\
& = & 1- \sum_{p,\nu} (1-\langle M_{ph\sigma}\rangle) |Y^\nu_{ph\sigma}|^2
\label{expOccupNumb-Hubb}
\end{eqnarray}
\noindent
where $\langle M_{ph\sigma}\rangle = \langle \tilde{n}_{h\sigma}\rangle + \langle \tilde{n}_{p\sigma}\rangle$, 
and  more complicated expressions for the quadratic terms which are similar to the preceding model cases. 
Those expressions are exactly the same as derived without the 
use of the vacuum in our earlier publication \cite{Jemai}.

\begin{figure}[ht]
  \begin{center}
    \leavevmode
     \epsfig{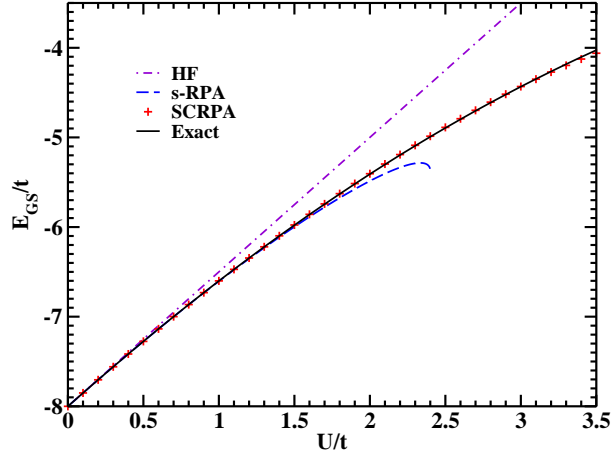}
      \caption{"(Color online)" 
      Energies of ground state in the standard RPA, SCRPA,
and exact cases as a function of $U/t$ for six sites with spin projection $m_{s} = 0$. 
}
  \label{EGS_Hubb}
 \end{center}
\end{figure}

In the Hubbard model, the results are again quite promising. We show in Fig.\ref{Rep2p3}, for a choice, the excitation spectrum for the momentum transfer $|q|=2\pi/3$. The results for $|q| = \pi/3, \pi$ are of similar quality \cite{Jemai}. In Fig.\ref{EGS_Hubb}, we show the ground state energy $E_{\mbox{GS}}= \langle Z|H|Z\rangle/\langle Z|Z\rangle $. There is good agreement with the exact solution and it presents a maximum error about $0.8$ percent at $U/t = 3.5$. 

Let us investigate the commutators $\langle [Q_{\nu}, Q^+_{\mu}]\rangle $ with $\mu \ne \nu$. Did we use the full expression (\ref{excOp}) for these operators, those operators would commute exactly for $\mu \ne \nu$. It is, therefore, interesting to estimate by how much this exact commutation rule is violated by our linearisation approximation of the operators. The expectation values of those commutators are given by the following expressions

\begin{eqnarray}
&&\langle\left[Q_{(q_2)\nu},Q^+_{(q_3)\mu}\right]\rangle  = 
\nonumber \\
&&~~~~~~~~ \sum_{\sigma} [ X^{(q_2)\nu}_{52\sigma}\langle S_{23\sigma} \rangle
                  -X^{(q_2)\nu}_{43\sigma}\langle S_{54\sigma} \rangle ]X^{(q_3)\mu}_{53\sigma}
\nonumber \\
&&~~~~~~ +\sum_{\sigma} [ Y^{(q_2)\nu}_{52\sigma}\langle S_{32\sigma} \rangle
                  -Y^{(q_2)\nu}_{43\sigma}\langle S_{45\sigma} \rangle ]Y^{(q_3)\mu}_{53\sigma} ~.  
               \label{commQnQm}
\end{eqnarray}

\begin{figure}[ht]
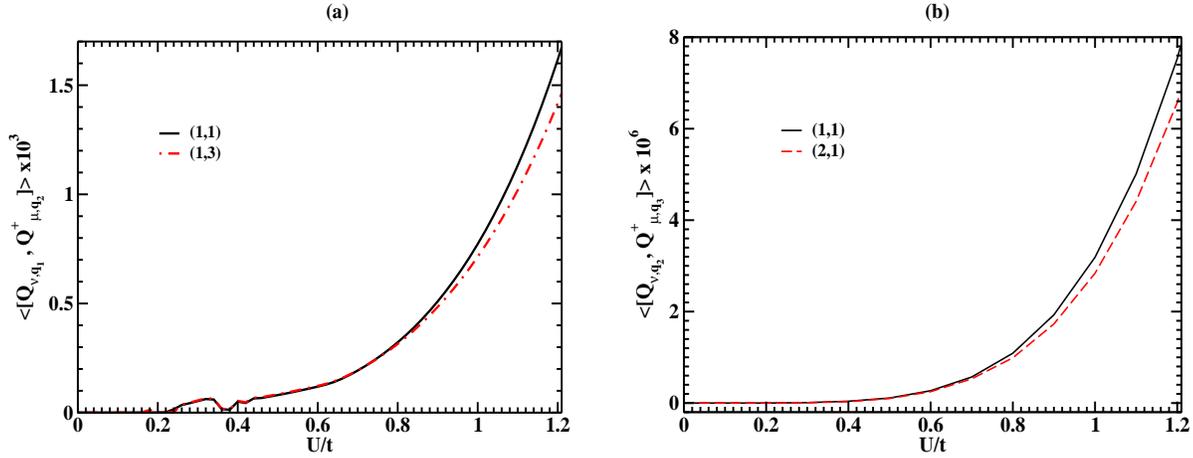

  \epsfig{figure=VQ1Q2-1-6.eps,width=7.5cm,height=6cm}
  \hspace{0.5cm}
  \epsfig{figure=VQ2Q3-12.eps,width=7.5cm,height=6cm}
\caption{"(Color online)"  
The commutator $\langle\left[Q_{\nu q_1},Q^+_{\mu q_2}\right]\rangle$ and $\langle\left[Q_{\nu,q_2},Q^+_{\mu,q_3}\right]\rangle$ as function of  $U/t$ with $q_1 =2\pi/3$, $q_2 =\pi$ and $(\nu,\mu)$ indicate the excited states in each channel. Only the worst cases are displayed. }
\label{VQ2Q3-12}
\end{figure}

\noindent
We can calculate the terms with $S_{ij}$,
\begin{eqnarray}
S_{h_1h_2\sigma} |Z\rangle \equiv b^{\dagger}_{h_1\sigma}b_{h_2\sigma}|Z\rangle = e^Z \bar S_{h_1h_2\sigma} |HF\rangle
\nonumber \\
S_{p_1p_2\sigma} |Z\rangle \equiv b^{\dagger}_{p_1\sigma}b_{p_2\sigma}|Z\rangle = e^Z \bar S_{p_1p_2\sigma} |HF\rangle
\end{eqnarray}
and
\begin{eqnarray}
\bar S_{h_1h_2\sigma} & = & e^{-Z} S_{h_1h_2\sigma} e^Z = S_{h_1h_2\sigma} +[S_{h_1h_2\sigma},Z] 
\nonumber \\
& = & S_{h_1h_2\sigma} + \sum_{ph,p'} z_{p'h_2,ph} J^+_{p'h_1\sigma} J^+_{ph,-\sigma} 
\nonumber \\
\bar S_{p_1p_2\sigma} & = & e^{-Z} S_{p_1p_2\sigma} e^Z = S_{p_1p_2\sigma} +[S_{p_1p_2\sigma},Z] 
\nonumber \\
& = & S_{p_1p_2\sigma} + \sum_{ph,h'} z_{p_2h',ph} J^+_{p_1h'\sigma} J^+_{ph,-\sigma} ~.
\end{eqnarray}

\noindent
Then,
\begin{eqnarray}
\langle S_{h_1h_2\sigma} \rangle & = & \sum_{ph,p'} z_{p'h_2,ph} \langle J^+_{p'h_1\sigma} J^+_{ph,-\sigma} \rangle
\nonumber \\
&=& \sum_{p'} \langle J^+_{p'h_1\sigma} J^-_{p'h_2,-\sigma} \rangle
\nonumber \\
\langle S_{p_1p_2\sigma}\rangle & = & \sum_{ph,h'} z_{p_2h',ph} \langle J^+_{p_1h'\sigma} J^+_{ph,-\sigma} \rangle 
\nonumber \\
&=& \sum_{h'} \langle J^+_{p_1h'\sigma} J^-_{p_2h',-\sigma} \rangle ~.
\label{ValSij} 
\end{eqnarray}
With these expressions, we can evaluate the commutators in (\ref{commQnQm}). The results are shown in 
Fig.\ref{VQ2Q3-12}. We see that the commutators are small of the order of 10$^{-3}$ for $\chi \sim 1$. 
We, therefore, can conclude that the RPA operators represent to very good approximation independent modes. 

\begin{figure}[ht]
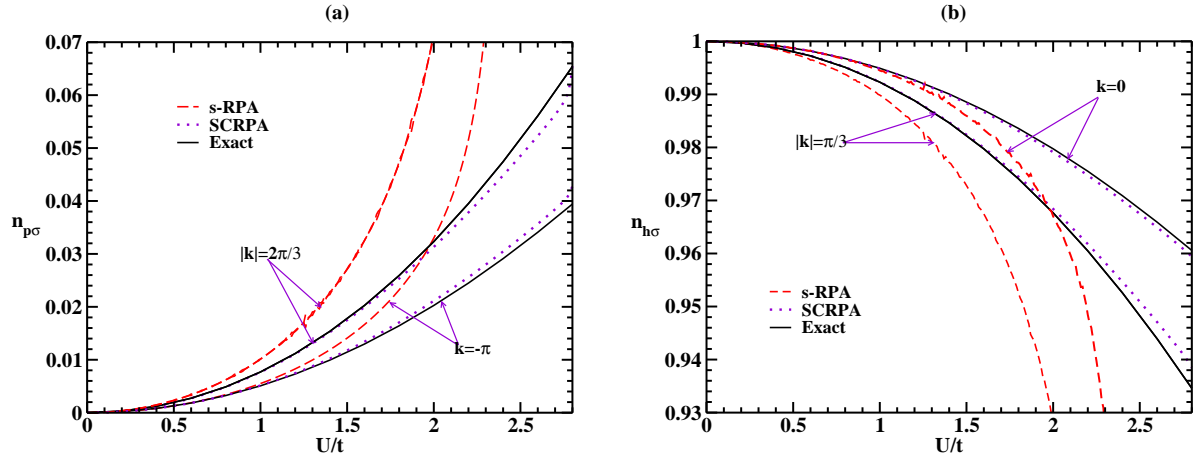

\begin{center}
  \epsfig{figure=n-p.eps,width=7.5cm,height=6cm}
  \hspace{0.5cm}
  \epsfig{figure=n-h.eps,width=7.5cm,height=6cm}
\caption{"(Color online)" 
Occupation numbers for particles and holes, $n_{p\sigma}$ and $n_{h\sigma}$, with SCRPA and exact solution as function of $U/t$. }
\label{n-ph}
\end{center}
\end{figure}

\noindent
In Fig.\ref{n-ph}, we show the occupation numbers, see Eq.(\ref{expOccupNumb-Hubb}). We see that with SCRPA they compare very well with the exact values and are very much improved over the corresponding values from standard RPA. It is worth noticing that particle number is conserved, i.e. what is depleted below the Fermi surface is exactly replaced by non-zero values above the Fermi surface. In the macroscopic limit, this would imply that the Luttinger theorem \cite{Luttinger} is respected.

\begin{figure}[ht]
\begin{center}
  \epsfig{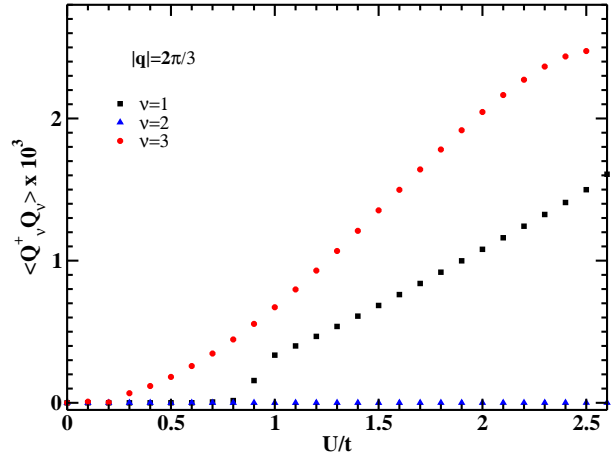}
\end{center}
\caption{"(Color online)" 
The mean value of $\langle Q^+_{\nu}Q_{\nu}\rangle $ for $|q|=2\pi/3$ and $\nu =1,2,3$ with SCRPA solution as function of $U/t$. Notice that the values for $\nu = 2$ are practically zero.}
\label{QPQ_NU}
\end{figure}

\noindent
In Fig.\ref{QPQ_NU}, we show the expectation value $\langle Q^+_{\nu} Q_{\nu}\rangle$ 
for $|q|= 2\pi/3$. In principle this expectation value will be zero, 
if the killing condition were fully satisfied. Because of our linearised form of the RPA 
operator this is violated. 
We, therefore, can see in Fig.\ref{QPQ_NU} where we present the worst cases that the error is of order $10^{-3}$. 
This should be compared to one, since it is part of the commutator which yields the normalisation. 
So, we find that the killing condition is only very slightly violated. Let us also remark that in more simplified models like the two level and three level Lipkin models these expectation values are exactly zero. Of course, this is only true within an expectation value.

\vspace{0.5cm}

\section{Summary, discussion, open problems}
\label{SummDisc}

In this paper we revised the SCRPA approach. An important formal step forward consists in the fact that the ground state wave function of the Coupled Cluster Theory (CCT)(truncated at the two body level) could be shown to be the vacuum to a generalised RPA operator. Though for the moment, 
we treated this generalisation only in an approximate way and, thus, 
stayed in the end with the usual linear form of the RPA operator,
the use of the CCT ground state wave function in the standard equation of motion method allowed to calculate all correlation functions in a natural way and to express them via the RPA amplitudes $X, Y$. This then leads to a kind of Hartree-Fock Bogoliubov, i.e., mean field theory for fermion pairs, that is, a self-consistent mean field approach for ph-(or pp-)modes. We revisited a number of exactly solvable many body models with this improved formalism. We calculated excitation energies, occupation numbers and other quantities which all are in quite good agreement with the exact results 
improving, for instance, the results around phase transition points. 
We showed that 
with the use of the standard RPA operator, the killing condition on the CCT vacuum is only slightly violated. The ensuing violation of the Pauli principle also remains very mild. Even in cases where it involves a maximum constraint like in the four particle case with single particle levels of only two-fold degeneracy, Pauli principle violations did not exceed a couple of percent. We, therefore, surmise that this is a general property of SCRPA. We will try in the future to apply SCRPA to realistic nuclear structure problems and compare the performance with, e.g., the results from shell model calculations.\\

We applied SCRPA only in the symmetry unbroken phases. We shall treat the symetry broken phases in a future work. In the past a less consistent version without the use of the CCT vacuum has been applied to the three level Lipkin model in a version with a continuously broken symmetry \cite{Delion}. There it had been shown that SCRPA in the 'deformed' region can be formulated in such a way that the Goldstone mode is reproduced what implies fullfillment of sum rules and conservation laws. This is a quite appealing  property, since usually it is very difficult to maintain this quality with numerically managable approaches which go beyond standard RPA. On the other hand, it, however, turned out in several cases \cite{Duke,Delion} that the transition from the 'spherical' regime to the 'deformed' one happens in a discontinuous way simulating a first order phase transition where there should not be any. We intend to investigate this problem in the future. Another point which shall be studied is implicit channel coupling. In principle channel coupling can occur in the example of the Hubbard model treated in Sect \ref{HubbardMod}. The total RPA matrix is block diagonal for each momentum transfer $q$ separately. However, via the nonlinearities all blocks can become coupled. We tried to include such couplings in our treatment but the results deteriorated. For the moment it is not clear whether this is a genuine effect or whether the formalism is not correctly generalised to this case. Further studies in this direction shall be undertaken. Since we are using the CCT vacuum, a natural question may be how SCRPA compares with CCT. In \cite{Dussel} such a comparison has been given. For the ground state energies the results are comparable. The advantage of SCRPA is that ground state and excitation energies are obtained in one shot from the same calculation. With CCT a separate calculation has to be performed to get to the excited state energies. Concluding we may say that, nonwithstanding some open problems, we think that substantial new insights and progress for a self consistent formulation of RPA could be achieved in this work. It is, indeed, our believ that this type of extension of RPA theory has good potential as a many body approach and that future developments and possibilites of that theory can further be exploited, see also \cite{Tohy}. For example, it could be interesting to include a generalised second RPA into the formalism, going beyond sub two CCT and introducing a still extended RPA operator \cite{Tohy}. Also generalisation to bosonic systems or mixtures could be an interesting task for future work.

\section {Acknowledgements}

The authors (specially P.S.) would like to acknowledge very fruitful collaboration in the past with J. Dukelsky on extensions of RPA theories. Extended discussions with M. Tohyama have also been appreciated.

\bibliographystyle{refer}
\bibliography{articles}


\end{document}